\documentclass{jfm}
\usepackage{graphicx}
\usepackage{epstopdf, epsfig}

\usepackage{amsmath,tikz,subcaption,verbatim,float}

\shorttitle{Shallow viscous current between walls}
\shortauthor{M-S. Liu and H. E. Huppert}

\title{Shallow current of viscous fluid flowing between diverging or converging walls}

\author{M-S. Liu\aff{1}$^,$\aff{2}
  \corresp{\email{msl63@cam.ac.uk}},
  H. E. Huppert\aff{3}$^,$\aff{4}}
\date{}
\affiliation{\aff{1}Homerton College, University of Cambridge, Cambridge CB2 8PH
\aff{2}Cavendish Laboratory, University of Cambridge, Cambridge CB3 0HE
\aff{3}King's College, University of Cambridge, Cambridge CB2 1ST
\aff{4}Department of Applied Mathematics and Theoretical Physics, University of Cambridge, Cambridge CB3 0WA}

\begin{document}

\maketitle

\begin{abstract}
We investigate the shallow flow of viscous fluid into and out of a channel whose gap width increases as a power-law ($x^n$), where $x$ is the downstream axis. The fluid flows slowly, while injected at a rate in the form of $t^\alpha$, where $t$ is time and $\alpha$ is a constant. The invading fluid has higher viscosity than the ambient fluid, thus avoiding Saffman-Taylor instability. Similarity solutions of the first kind for the outflow problem are found using approximations of lubrication theory. Zheng et al [2014] studied the deep-channel case and found divergent behaviour of the similarity variable as $n\rightarrow1$ and $n\rightarrow3$, when fluid flows into and out of the channel respectively. No divergence is found in the shallow case presented here. The characteristic equilibration time for the numerically simulated constant-volume flow to converge to the similarity solution is calculated assuming inverse dependence on the ratio disagreement between the current front using the method of lines (MOL). The inverse power dependence between equilibration time and ratio disagreement is found for channels of different powers. A similarity solution of the second kind for the inflow problem is found using the phase plane formalism and the bisection method. An exponential decay relationship is found between $n$ and the degree $\delta$ of the similarity variable $xt^{-\delta}$, which does not show any divergent behaviour for large $n$. An asymptotic behaviour is found for $\delta$ that approaches $1/2$ as $n\rightarrow\infty$.
\end{abstract}


\section{Introduction}

There are a lot of similarities between the journey of a glacier and the flow of golden syrup if we record and speed up the glacier's flowing on a camera. If the glacier flows without any disturbance, e.g., seasonal changes, we might even find it to be identical to the flow of golden syrup in a gap, up to scaling transformations. The shape of a current at different points in time are the same up to scaling when it reaches a steady state when memories of the initial releasing conditions are lost. This idea of scaling symmetry is a powerful tool that finds particular solutions for partial differential equations, called similarity solutions (Barenblatt [1996]). Notably, the self-similarity method solves numerous problems in the field of gravity currents. Huppert [1982] modelled the two-dimensional and axisymmetric spread of shallow and viscous currents using the similarity method, which agree with the experimental data. The method of similarity calculates the behaviour of the current using scaling symmetry. The solution derived represents the current in a stationary condition when the memory of the initial condition is lost, which is unrealistic considering the current must differ depending on the way it is released. Ball \& Huppert [2019] and Webber \& Huppert [2019] discuss the  timescale at which an axisymmetric current converges to the self-similar solution. A similar construction is used herein to investigate the timescale of real flow approaching the similarity solutions, and we found different leading power-terms from the axisymmetric case studied by Ball and Webber \& Huppert. 

This paper focuses on the viscous spreading of gravity currents in a power-law channel, where the spreading is slow enough for the inertial force to be negligible, and the height of the fluid is shallow, so the approximations of lubrication theory apply. We use the self-similar method to find the similarity solution of the first kind for the viscous current flowing out of the channel and find the characteristic time of the spread using dimensional analysis and the method of lines (MOL) numerical scheme as employed in \verb|Mathematica|. We study the inflow problem using a similar method, except without the global conservation condition. Having one less govering equation but the same degrees of freedom means self-similar solutions of the first kind do not exist. Gratton \& Minotti [1990] developed a phase-plane formalism that we adapt to find self-similar parameters for different power-law channels, showing the existence of self-similar solutions of the second-kind.

The self-similar parameters contain information about the shape of the current, \textit{i.e.}, if two flows share the same parameter, their shapes agree up to a scaling constant. Using computational analysis, we find that the self-similar parameters of the inflow problem approach $1/2$ asymptotically through exponential decay. The asymptotic behaviour means the shape of the flow stablises as it evolves, and tends to a constant shape at infinity.

\section{Theory}
\subsection{Outflow from the origin}

Consider fluid of density $\rho$ released from the origin of a channel whose width increases in the polynomial form of $b(x)=b_0x^n$, where $b_0$ is a fixed constant. The space is filled with ambient fluid of density $\rho-\Delta \rho$ ($\Delta \rho>0$) with lower viscosity than the invading fluid, so Saffman-Taylor instability does not occur. If the height of the fluid is significantly smaller than the width, i.e., relatively shallow, the viscous force exerted by the bottom plate dictates the resistance to the flow. The bottom-plate dissipation of one dimensional and axisymmetric gravity currents have been studied by Huppert [1982], using the approximations of lubrication theory, supposing zero shear stress at the top of the gravity current. We will follow the same line of logic to derive an approximation to the Stokes equation, together with a different continuity equation that satisfies the streamwise heterogeneity condition.

Assume the current flows slowly, so the fluid is instantaneously hydrostatic. The pressure is then given by $p(x,z) = p_0+\rho g(h-z)$, where $p_0$ is some constant, $g$ is gravity and the $z$ axis, is vertically upwards with $z=0$ at the base. The viscous force balances with the pressure gradient, leading to
\begin{equation}
    \frac{1}{\rho}\frac{\partial p}{\partial x}=\nu\frac{\partial^2u}{\partial z^2} = g'\frac{\partial h}{\partial x},
    \label{eq:firststokes}
\end{equation}
where $g'=(\Delta \rho/\rho) g$ and $x$ is along the channel. The fluid travels with velocity $u$ in the $x$ direction. Equation \eqref{eq:firststokes} is an approximation to the Stokes equation, when the velocity variation in $z$ dominates, which is the Navier-Stokes equation when the inertial terms are negligible. Applying the non-slip condition on the bottom plate ($u|_{h=0}$) and the continuity of shear stress ($\partial u/\partial z|_{z=h_{\pm}}=0$ if the ambient is a lot less viscous, e.g. honey intruding air), we obtain
\begin{equation}
    u=-\frac{g'}{2\nu}\frac{\partial h}{\partial x}z(2h-z).
\end{equation}
Averaging the velocity in the $z$ direction, we obtain one of the two governing equations
\begin{equation}
    \bar{u}=-\frac{\Delta \rho g}{3\mu}h^2\frac{\partial h}{\partial x},
    \label{eq:secondstokes}
\end{equation}
where $\bar{u}$ is the streamwise velocity averaged in the $z$ direction.
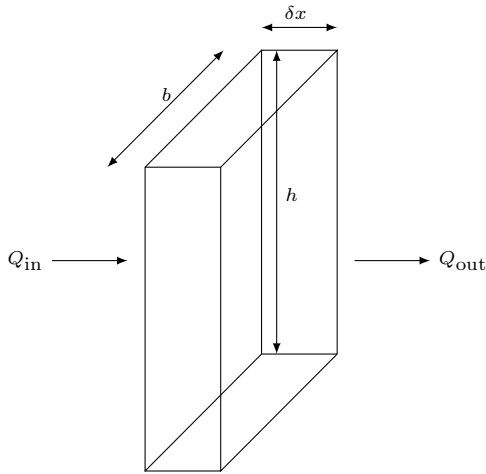
\begin{figure}
    \centering
    \begin{tikzpicture}[>=latex, font=\scriptsize]
        \newcommand{\x}{1}
        \newcommand{\xx}{\x/2}

        \draw (0,0,0) -- (0,4,0) -- (0,4,4) -- (0,0,4) -- (0,0,0);
        \draw (\x,0,0) -- (\x,4,0) -- (\x,4,4) -- (\x,0,4) -- (\x,0,0);
        \draw [<-] (-.5,4,0) -- node [left] {$b$} (-.5,4,3);
        \draw [->] (-.5,4,3) -- (-.5,4,4);
        \draw [<-] (0,4.3,0)-- node [above] {$\delta x$} (0.9,4.3,0);
        \draw [->] (0.9,4.3,0)--(\x,4.3,0);
        \draw [<-] (0.2,4,0) -- node [right] {$h$} (0.2,0.2,0);
        \draw [->] (0.2,0.2,0) -- (0.2,0,0);

        \draw (0,0,0)--(\x,0,0);
        \draw (0,4,0)--(\x,4,0);
        \draw (0,0,4)--(\x,0,4);
        \draw (0,4,4)--(\x,4,4);

        \draw [->] (-2*\x,2,2) node [left] {$Q_{\mbox{in}}$} -- (-\x,2,2);
        \draw [->] (2*\x,2,2) -- (3*\x,2,2) node [right] {$Q_{\mbox{out}}$};
    \end{tikzpicture}
\caption{A diagram showing fluid entering and leaving a thin rectangular region.}
\label{fig:flow}
\end{figure}

\begin{figure}
    \centering
    \includegraphics[width=0.6\linewidth]{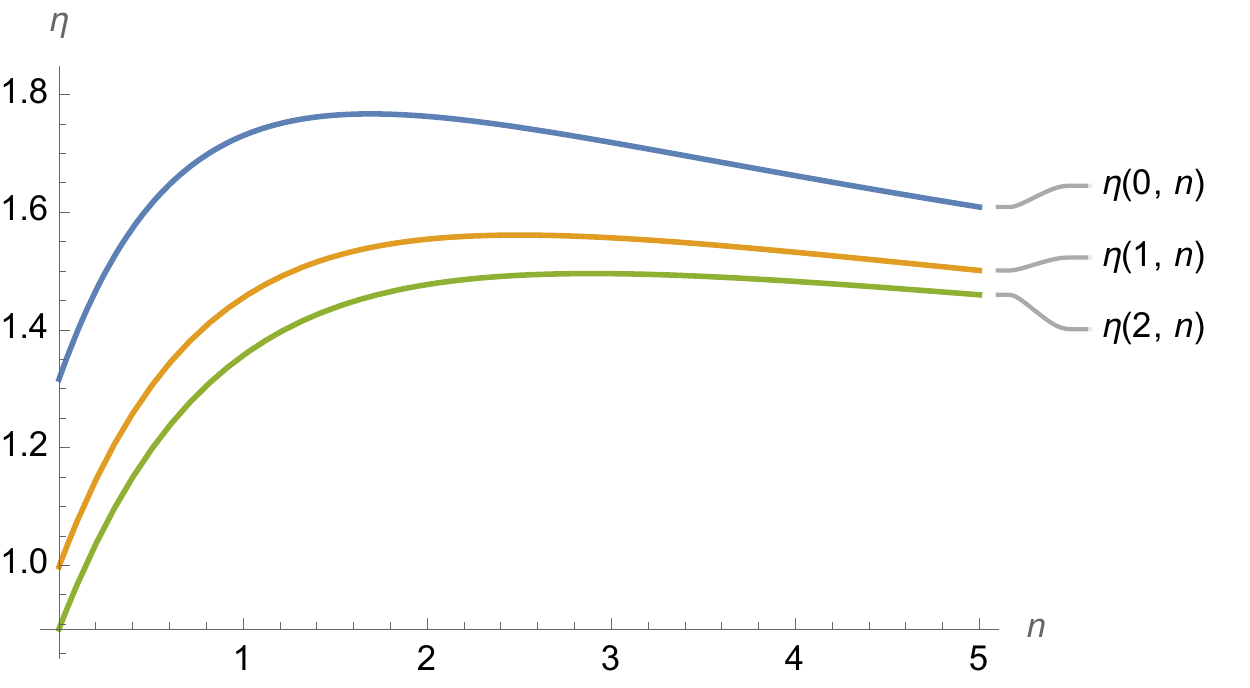}
    \caption{A plot of $\eta$ against $n$ derived using the similarity method of the first kind, $\eta_f(\alpha,n)$ is the unique constant of proportionality for a flow of injection power $\alpha$ in an $n^\text{th}$ power channel. The graph shows the behaviours for constant volume $\alpha=0$, constant injection rate $\alpha=1$, and $\alpha=2$ in different channels.}
    \label{fig:etaf}
\end{figure}

The incompressibility conditions suggests that whatever enters a box of width $\delta x$ as shown in figure \ref{fig:flow} must either exit or be compensated by a change of height, which leads to

\begin{equation}
    \frac{\partial Q}{\partial x} = -\frac{\partial (hb)}{\partial t},
\end{equation}
where the flux $Q$ is the product of averaged velocity $\bar{u}$ and the cross-sectional area $h\cdot b$. The continuity equation is therefore
\begin{equation}
    \label{eq:continuity}
    \frac{\partial h}{\partial t} +\frac{1}{x^n}\frac{\partial}{\partial x}\big[(b_0x^n)h\bar{u}\big] = 0.
\end{equation}
Together with \eqref{eq:secondstokes}, we derive the nonlinear partial differential equation that governs the height change with $x$ and $t$,
\begin{equation}
    \label{eq:height}
    \frac{\partial h}{\partial t} -\frac{\beta}{x^n}\frac{\partial}{\partial x}(x^nh^3\frac{\partial h}{\partial x}) = 0,
\end{equation}
where $\beta=\Delta \rho g/3\mu$. The overall volume of the fluid is conserved and equal to the rate of injection, which we assume to take the general power-law form $t^\alpha$. Hence
\begin{equation}
    \int_0^{x_f}hx^ndx=B t^\alpha,
    \label{eq:conservation}
\end{equation}
where $B$ is the proportionality constant. Equations \eqref{eq:height} and \eqref{eq:conservation}, together with the current front condition $h[x_f(t)]=0$, contain sufficient information to determine $h(x,t)$. Assuming $h(x,t)$ exists in an intermediate asymptotic regime where the solutions are self-similar, we can find the similarity solution of the first kind using scaling analysis (Barenblatt [1966]). The similarity variable is found to be
\begin{equation}
    \eta=x(\beta B^3)^{\frac{-1}{3n+5}}t^{-\frac{3\alpha+1}{3n+5}}.
    \label{eq:simvar}
\end{equation}
We define $\eta_f$ to be the value of $\eta$ at $x_f(t)$, which is the position of the current front. Thus from \eqref{eq:simvar}
\begin{equation}
    x_f(t)=\eta_f(\beta B^3)^{\frac{1}{3n+5}}t^{\frac{3\alpha+1}{3n+5}}
    \label{eq:fluidfront}
\end{equation}
and we can determine the similarity solution of $h$ in terms of $\eta$ as
\begin{equation}
    h=\eta_f^{\frac{2}{3}}\beta^{-\frac{n+1}{3n+5}}B^{\frac{2}{3n+5}}t^{\frac{2\alpha-n-1}{3n+5}}\phi(y),
    \label{eq:heightfirst}
\end{equation}
where $y=\eta/\eta_f=x/x_f$. Substituting \eqref{eq:heightfirst} into \eqref{eq:height} and \eqref{eq:conservation}, we find the following ordinary differential equation for $\phi$ and expression for $\eta_f$
\begin{subequations}
    \label{eq:outfloweq}
    \begin{equation}
        \label{subeq:phiode}
        y(\phi^3\phi')'+n\phi^3\phi'+\frac{3\alpha+1}{3n+5}y^2\phi'-\frac{2\alpha-n-1}{3n+5}y\phi=0,
    \end{equation}
and
    \begin{equation}
        \label{subeq:conserve}
        \eta_f=\Big(\int_0^1y^n\phi dy\Big)^{-\frac{3}{3n+5}}.
    \end{equation}
\end{subequations}
The front of the current corresponds to $y=1$, so the boundary condition relevant for \eqref{subeq:phiode} is $\phi(1)=0$. Expanding $\phi$ about $y=1$, we obtain the leading terms of $\phi$ as
\begin{equation}
    \phi(y)=\Big[3\Big(\frac{3\alpha+1}{3n+5}\Big)\Big]^{1/3}(1-y)^{1/3}\Big[1+\frac{1}{32}\frac{9(n+1)\alpha-2}{3\alpha+1}(1-y)+\mathcal{O}(1-y)^2\Big].
\end{equation}
Therefore, when $y\rightarrow 1$, the similarity solution approximates to
\begin{equation}
    h\sim\Big[3\Big(\frac{3\alpha+1}{3n+5}\Big)\Big]^{\frac{1}{3}}\eta_f^{\frac{2}{3}}\beta^{-\frac{n+1}{3n+5}}B^{\frac{2}{3n+5}}(1-y)^\frac{1}{3}t^{\frac{2\alpha-n-1}{3n+5}},
    \label{eq:firstorderh}
\end{equation}
and $\eta_f$ can be evaluated as
\begin{equation}
    \eta_f(\alpha,n)=\bigg\{\Big(\frac{9\alpha+3}{3n+5}\Big)\Big[\frac{\Gamma(4/3)\Gamma(n+1)}{\Gamma(n+7/3)}\Big]^3\bigg\}^\frac{-1}{3n+5},
\end{equation}
where $\Gamma(z)=\int_0^\infty x^{z-1}e^{-x}dx$ is the standard gamma function. Figure \ref{fig:etaf} shows the shape of $\eta_f$ for $\alpha=0,1,2$ respectively.

Hence, from integrating \eqref{subeq:conserve}, $\eta_f$ is a constant that characterises the shape of the channel and the nature of the flow, e.g., for  $\alpha=0$ and $n=1$
\begin{equation}
    \eta_f(0,1)= \Big(\frac{3^7}{2^9\times 7^3}\Big)^{-1/8}\approx 1.73\dots.
    \label{eq:onepoint}
\end{equation}
The similarity solution we have obtained from dimensional analysis assumes the flow is self-similar, i.e., the shape of the current at different times are related through a scaling transformation. However, the release of the currents is not perfect in real-world experiments, and the real flow differs from the self-similar flow. To investigate how quickly the real flow converges to the self-similar flow as the memory of the initial releasing condition is lost, we investigate the characteristic equilibration time $\tau$, which is the time it takes for the real and self-similar flow to agree to a certain extent. The exact form of $\tau$ will be different for different initial conditions. To first-degree approximation (whose order we shall determine), this characteristic time is determined by the physical variables $\beta$ and $B$, and some variable that parametrises the initial condition.

One way of parameterising the initial configuration of the fluid is by considering the ratio between the height at the origin and the extent of the flow, i.e., through the aspect ratio $\gamma = h(0,t)/r_f(t)$. Using \eqref{eq:fluidfront} and \eqref{eq:firstorderh}, we obtain
\begin{equation}
    \gamma=\Big[3\Big(\frac{3\alpha+1}{3n+5}\Big)\Big]^{1/3}\eta_f^{-1/3}\beta^{-\frac{n+2}{3n+5}}B^{-\frac{1}{3n+5}}t^{-\frac{\alpha+n+2}{3n+5}}.
\end{equation}
Specifically, for a fixed-volume ($\alpha=0$) fluid flowing in a linear gap ($n=1$), we can calculate the value of $\eta_f$ from \eqref{eq:onepoint} and find the dependence between the equilibration time and initial aspect ratio through dimensional analysis. We can imagine setting up a gate at $x_0$ confining fluid of constant height $h_0$ as the simplest initial condition. Setting $\alpha=0$ in equation \eqref{eq:conservation}, we obtain
\begin{equation}
    B=\int_0^{x_0} h_0 x^n dx,
\end{equation}
 hence $B\propto h_0 x_0^{n+1} = \gamma_0 x_0^{n+2}$. Substituting back to \eqref{eq:fluidfront}, bearing in mind that $\eta_f$ is a constant for fixed $\alpha$ and $n$, we determine the equilibration time satisfying
\begin{equation}
    B^{\frac{1}{n+2}}\beta \tau \gamma_0^{\frac{3n+5}{n+2}}=F(p,\mbox{ shape}),
\end{equation}
where $p$ is the disagreement ratio between the real-flow front ($x_r$) and the front of the similarity solution ($x_s$), i.e.,
\begin{equation}
    p=\frac{|x_r-x_s|}{x_r},
\end{equation}
We now consider what happens at extreme values of $p$. As $p$ approaches $0$, the disagreement between the similarity solution and the real solution also approaches zero, which intuitively takes forever to achieve, i.e., $\tau\rightarrow\infty$. However, there could be a huge disparity ($p\rightarrow \infty$) between the two solutions when the fluid is initially released ($\tau\rightarrow 0$). With these two intuitive observations in mind, we can guess that $p$ decreases with $\tau$ and diverges as $p\rightarrow 0^+$. The exact analytic dependence is not straightforward, but Ball \& Huppert (2019) and Webber \& Huppert (2019) showed that the inversely proportional relationship ($\tau\propto p^{-1}$) works as an excellent first-order approximation for the radially symmetric case. Although the exact proportionality for the power-law channel might differ from the radially symmetric case, we can assume that
\begin{equation}
    B^{\frac{1}{n+2}}\beta \tau \gamma_0^{\frac{3n+5}{n+2}}=p^{- \chi(n)}F(p,\mbox{ shape}),
\end{equation}
where $\chi=\chi (n)>0$ depends on the power of the channel. $\chi$ must be positive for the solution to be physical, \textit{i.e.}, the solutions converge as time goes on.

\begin{figure}
    \centering
    \includegraphics[width=0.6\linewidth]{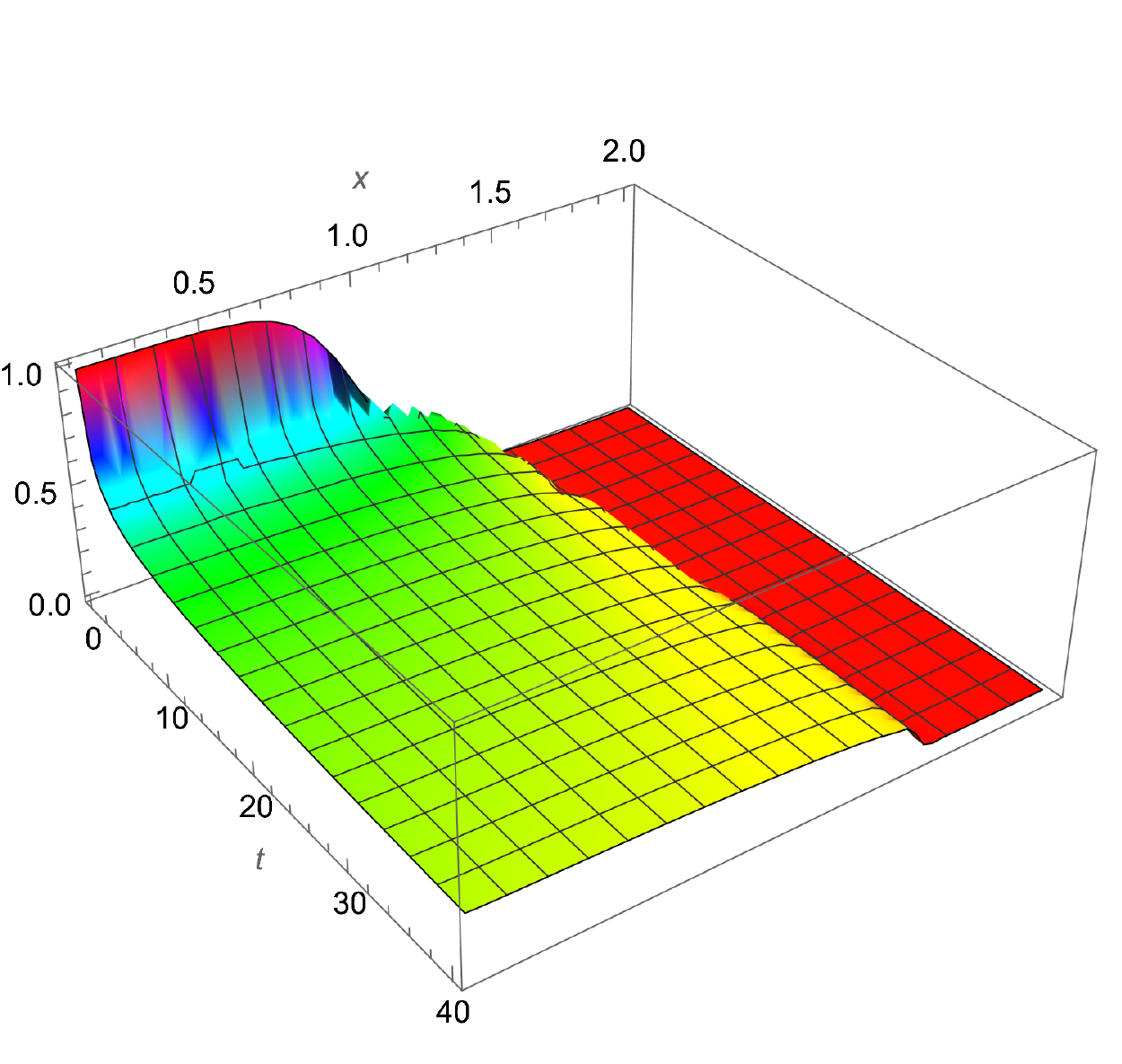}
    \caption{The evolution of the current across a period of time solved numerically, in the case of $\alpha=0,n=2$, where the colour signifies the height of the fluid. }
    \label{fig:n2demo}
\end{figure}

\begin{figure}
    \centering
    \includegraphics[width=0.6\linewidth]{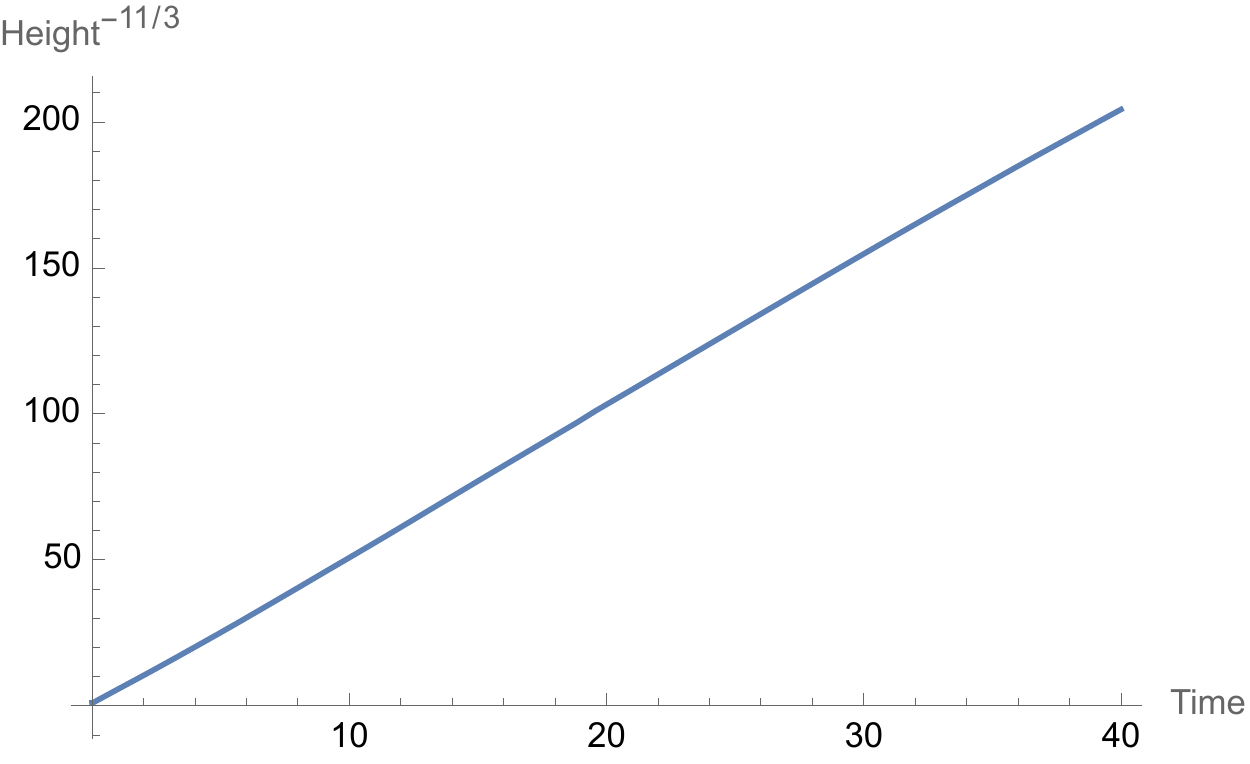}
    \caption{To check the validity of the simulation, we fix $y$ in equation \eqref{eq:firstorderh} and show the predicted proportionality explicitly.}
    \label{fig:simConfirmation}
\end{figure}

We can find $\chi(n)$ computationally by simulating the flow of the current using \verb|NDSolve| in \verb|Mathematica|, and simplifying the boundary condition \eqref{eq:conservation} using the method shown in appendix \ref{appendix:bc}. The first step is to find the numerical solution, which is shown in figure \ref{fig:n2demo} for the example of $\alpha=0$ (constant volume), $n=2$ flow. To demonstrate the validity of the simulation, we pick and fix $y$ from equation \eqref{eq:firstorderh} and show the proportionality predicted by the similarity solution by plotting $h^{-11/2},\alpha=0,n=2$ against $t$ as shown in figure \ref{fig:simConfirmation}.

\begin{figure}
    \centering
    \includegraphics[width=0.6\linewidth]{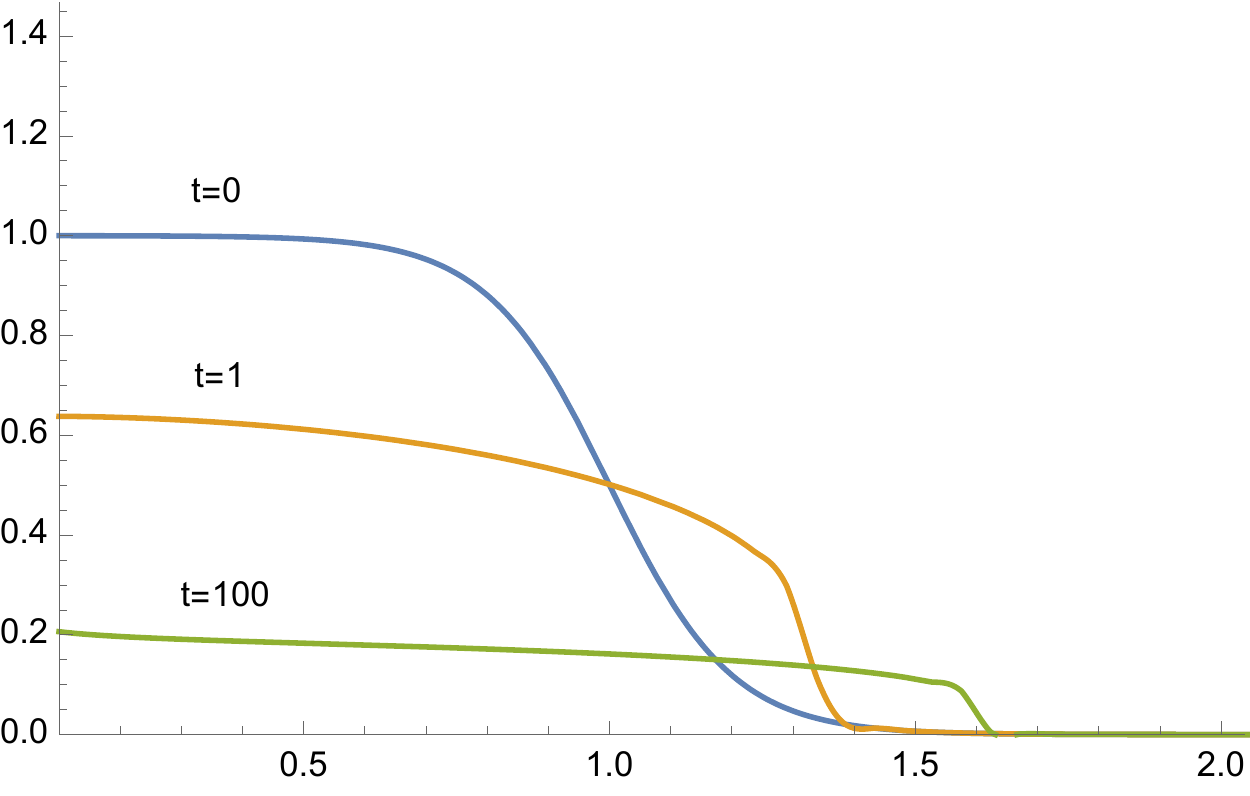}
    \caption{Fluid profile at $t=0,1,100$ for $\alpha=0,n=2$}
    \label{fig:fluidEvo}
\end{figure}

\begin{figure}
     \centering
     \begin{subfigure}{0.45\textwidth}
         \centering
         \includegraphics[width=\linewidth]{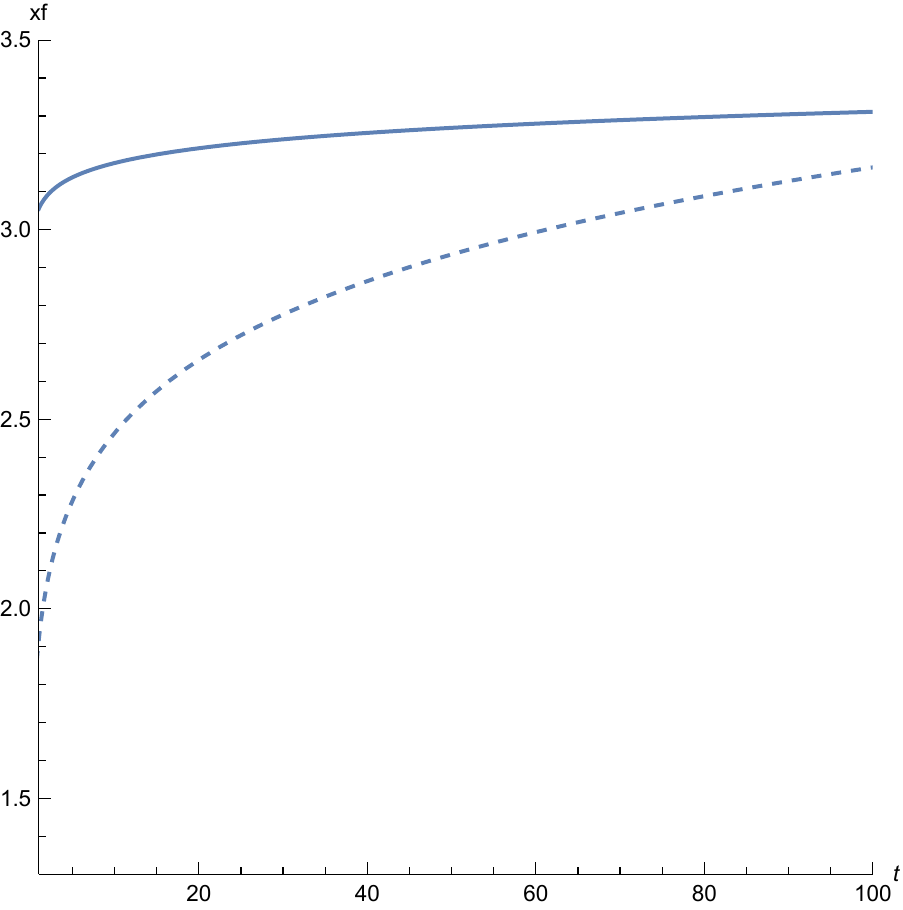}
         \caption{Simulation and the similarity solution}
         \label{fig:svss}
     \end{subfigure}%
     \begin{subfigure}{0.45\textwidth}
         \centering
         \includegraphics[width=\linewidth]{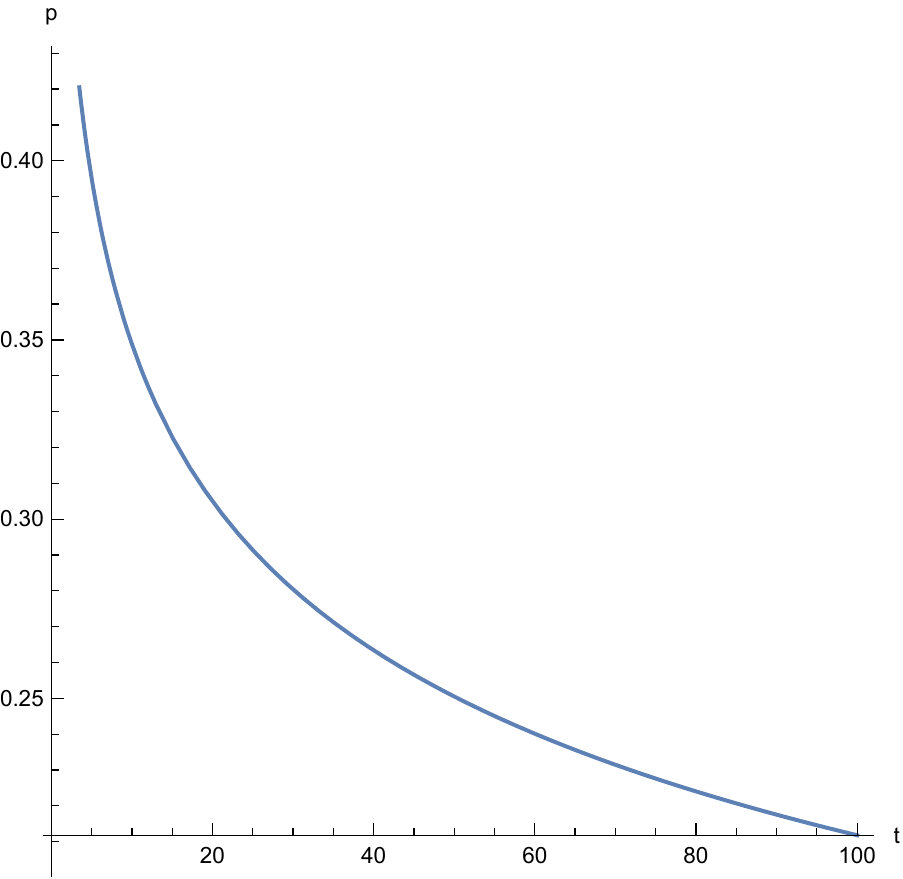}
         \caption{$p$ against $t$}
         \label{fig:pplot}
     \end{subfigure}
     \caption{\subref{fig:svss} Comparison between the simulation and the similarity solution for $\alpha=0,n=2$, where the solid line represents the simulation and dashed line represents the similarity solution \subref{fig:pplot} The ratio difference $p$ plotted against time}
    \label{fig:xfDif}
\end{figure}

We can then find the timescale of asymptotpic approach to the similarity solution by plotting the difference ratio ($p$) against $t$ for each $n$. An important point to make is that we have chosen a smooth profile as the initial shape of the fluid to improve the accuracy of the discretisation, as shown in figure \ref{fig:fluidEvo}. This comes at the price of a less well-defined fluid front $x_f$, which is closely approximated as the point of inflection of the fluid profile at fixed $t$. The offset also does not affect the value of $\chi(n)$ that we care more about. Figure \ref{fig:xfDif} shows the difference between the simulated moving front and that predicted by the similarity solution. Figure \ref{fig:pplot} shows a decaying trend that we can use to find $\chi(n)$ with the \verb|FindFit| function in \verb|Mathematica|.

Iterating the same process for different values of $n$, we can find $\chi(n)$ numerically. The result is presented in figure \ref{fig:chiofn}, and a table of simulated results for integer-power channels is presented in appendix \ref{appendix:table}. The idea is that when performing experiments, one can measure the difference between the real flow in experiment and the similarity solution empirically, and anticipate the difference to drop by $p\propto \tau^{-\chi(n)}$ over time $\tau$. Furthermore, we observe a trend of $\chi\propto 1/n$ as shown by the plot fit in figure \ref{fig:chiofn}.
\begin{figure}
    \centering
    \includegraphics[width=0.6\linewidth]{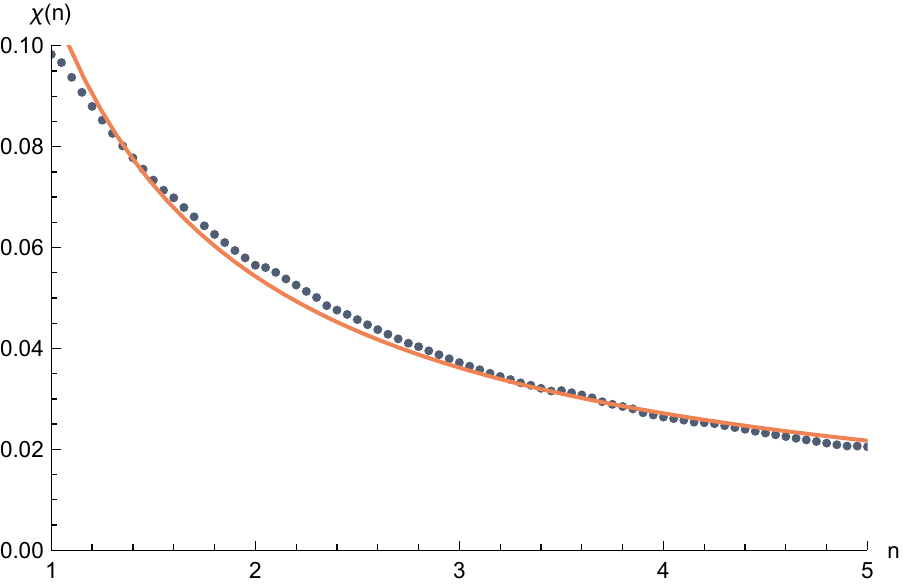}
    \caption{$\chi(n)$ plotted against $n$, as calculated computationally. The solid orange line represents $\sim 1/n$ up to scaling constant, and shows close agreement with the simulated result, indicating possible inverse proportionality.}
    \label{fig:chiofn}
\end{figure}
\subsection{Inflow towards the origin}
Consider instead the current flowing towards the origin. How does this change the equation of motion? Equation \eqref{eq:height} is invariant under this transformation since the geometry of the flow, including the channel, is locally the same as before, and the approximation to the Stokes equation still holds. However, the conservation equation \eqref{eq:conservation} fails because we do not integrate from the origin but back from the current front to infinity instead. In experiments, the domain of integration terminates at a finite distance, ideally far enough for the memory of the initial releasing conditions to be lost, so self-similarity arises. Losing one of the boundary conditions means the solution to $h(x,t)$ is not unique, and the problem becomes more difficult. The governing equations are the Stokes equation \eqref{eq:secondstokes} and the continuity equation \eqref{eq:continuity} as before
\begin{subequations}
    \label{eq:GovEq}
    \begin{equation}
        \label{subeq:GovEq1}
        \frac{\partial h}{\partial t} +\frac{1}{x^n}\frac{\partial}{\partial x}(x^nhu) = 0,
    \end{equation}
    \begin{equation}
        \label{subeq:GovEq2}
        u=-\beta h^2\frac{\partial h}{\partial x},
    \end{equation}
\end{subequations}
where instead of substituting $u$ to form \eqref{eq:height}, we kept them as they were.

Gratton and Minotti [1990] studied a similar set of equations using the phase-plane formalism, which we use to further investigate the nature of inflow gravity current. Using scaling analysis, we arrive at
\begin{subequations}
    \begin{equation}
        \label{subeq:gova}
        u(x,t)=xU(x,t)/t,
    \end{equation}
    \begin{equation}
        \label{eq-b}
        h(x,t) = \Big[x^2H(x,t)/\beta t\Big]^{1/3},
    \end{equation}
\end{subequations}
where both $U(x,t)$ and $H(x,t)$ are dimensionless. Substituting these representations into \eqref{eq:GovEq}, we obtain
\begin{subequations}
    \label{eq:GovEqDless}
    \begin{equation}
        2H+3U+x\frac{\partial H}{\partial x} = 0,
        \label{eq:GovEqDlessa}
    \end{equation}
    \begin{equation}
        H-t\frac{\partial H}{\partial t}-(3n+5)UH-x(3H\frac{\partial U}{\partial x}+U\frac{\partial H}{\partial x})=0,
    \end{equation}
\end{subequations}
which are a set of coupled nonlinear partial differential equations (PDE). Most PDE's do not have analytic solutions. However, we expect the currents to have some degree of self-similarity as they propagate, i.e., the currents only differ by a similarity transform across time. We can exploit this by defining similarity variables and reducing the number of independent variables. Specifically, the similarity condition is only satisfied when the current is far enough from the release source, but has not quite reached the origin, i.e., in an \textit{intermediate} stage (Barenblatt [2003]). There are two types of similarity solutions. The degree of self-similarity of a system depends on the geometry of the flow. With complete self-similarity, we can derive a full analytic description using scaling arguments as presented above; however, numerical analysis is required if the similarity is incomplete.

To eliminate one independent variable, we define a similarity variable $\eta=xt^{-\delta}$ and substitute this into equations \eqref{eq:GovEqDless}. Eliminating $\eta$ and rewriting equtaion \eqref{subeq:gova}, we obtain
\begin{subequations}
    \label{eq:PhasePlane}
    \begin{equation}
        \label{subeq:auto}
        \frac{dU}{dH} = \frac{H[3(n+1)U+2\delta-1]+3U(\delta-U)}{3H(3U+2H)},
    \end{equation}
    \begin{equation}
        \label{subeq:eta}
        \frac{d\ln|\eta|}{dH}=-\frac{1}{2H+3U}.
    \end{equation}
\end{subequations}
Equation \eqref{subeq:auto} is an autonomous equation for $U$ and $H$, which can be solved analytically in special cases and numerically in most cases. However, it is crucial to identify the initial conditions before performing the integration. The path along which we integrate is guided by the phase plane vector field, and the endpoints coincide with critical points that decide the boundary conditions and the shape of the current. Once the integral path $U(H)$ has been found, equation \eqref{subeq:eta} can be integrated to find $\eta(H)$, with $U(\eta)$ and $H(\eta)$  then found through inversion.

The critical points are locally stationary, which can be found by setting both the denominator and numerator to zero in \eqref{subeq:auto}. There are three finite and three infinite critical points, each representing a different initial condition. The points at infinity represent different types of boundary conditions, including the flow of moving sinks, which is treated in Gratton and Minotti [1990], but unrelated to the inflow problem at hand. The finite critical points are

\begin{enumerate}
    \item $O:(H,U)=(0,0)$, the fluid is stationary and has constant height,
    \item $A:(H,U)=(0,\delta)$, the current height is zero at the front and travels at a finite velocity, i.e., the advancing front of the viscous gravity current,
    \item $B:(H,U)=[-3/2(5+3n),1/(5+3n)]$, the current height and velocity $h\propto (-x^2/t)^{1/3}$ and $u\propto x/t$ representing a flow outwards from the channel. Integration paths around point $B$ spiral endlessly, and so $U$ and $H$ exhibit oscillatory behaviour not found in the physical variables $u$ and $h$.
\end{enumerate}
The flow to the origin is represented by the integral path from A to O, which only exists under specific values of $\delta(n)$. By finding $\delta(n)$, we can demonstrate the existence of a similarity solution of the second kind, and the actual flow can be simulated numerically using the phase plane.

The value of $\delta(n)$ can be found computationally by altering $n$ and changing the value of $\delta$ until the integral path shooting from perturbation at A reaches O (Zheng et al [2014] and [2022]). The generating perturbation to first order around point A can be found by linearising \eqref{subeq:auto}, and the eigenvectors indicate which integral path passes through A. Let  $\mathbf{r}_A\rightarrow \mathbf{r}_A + \delta\mathbf{r}$, where $\delta\mathbf{r}=(\eta,\mu)$ and consider up to the first order,
\begin{equation}
    \frac{dU}{dH}\Big|_A = \frac{\eta[3(1+n)(\delta +\mu)+2\delta -1]-3\mu(\delta+\mu)}{3\eta(2\eta+3\delta+3\mu)}\approx \frac{[(3n+5)\delta-1]\eta-3\mu\delta}{9\eta\delta},
\end{equation}
\begin{equation}
    \delta\mathbf{r}\approx
    \begin{pmatrix}
        9\delta&0\\(5+3n)\delta-1&-3\delta
    \end{pmatrix}
    \mathbf{r}.
\end{equation}
The eigenvectors corresponding to the linearised matrix are
\begin{align}
    \lambda_1&=9\delta, &\mathbf{e}_1&= [12\delta,(5+3n)\delta-1];\\
    \lambda_2&=-3\delta, &\mathbf{e}_2&=(0,1).
\end{align}
\begin{figure}
    \centering
    \includegraphics[width=\textwidth]{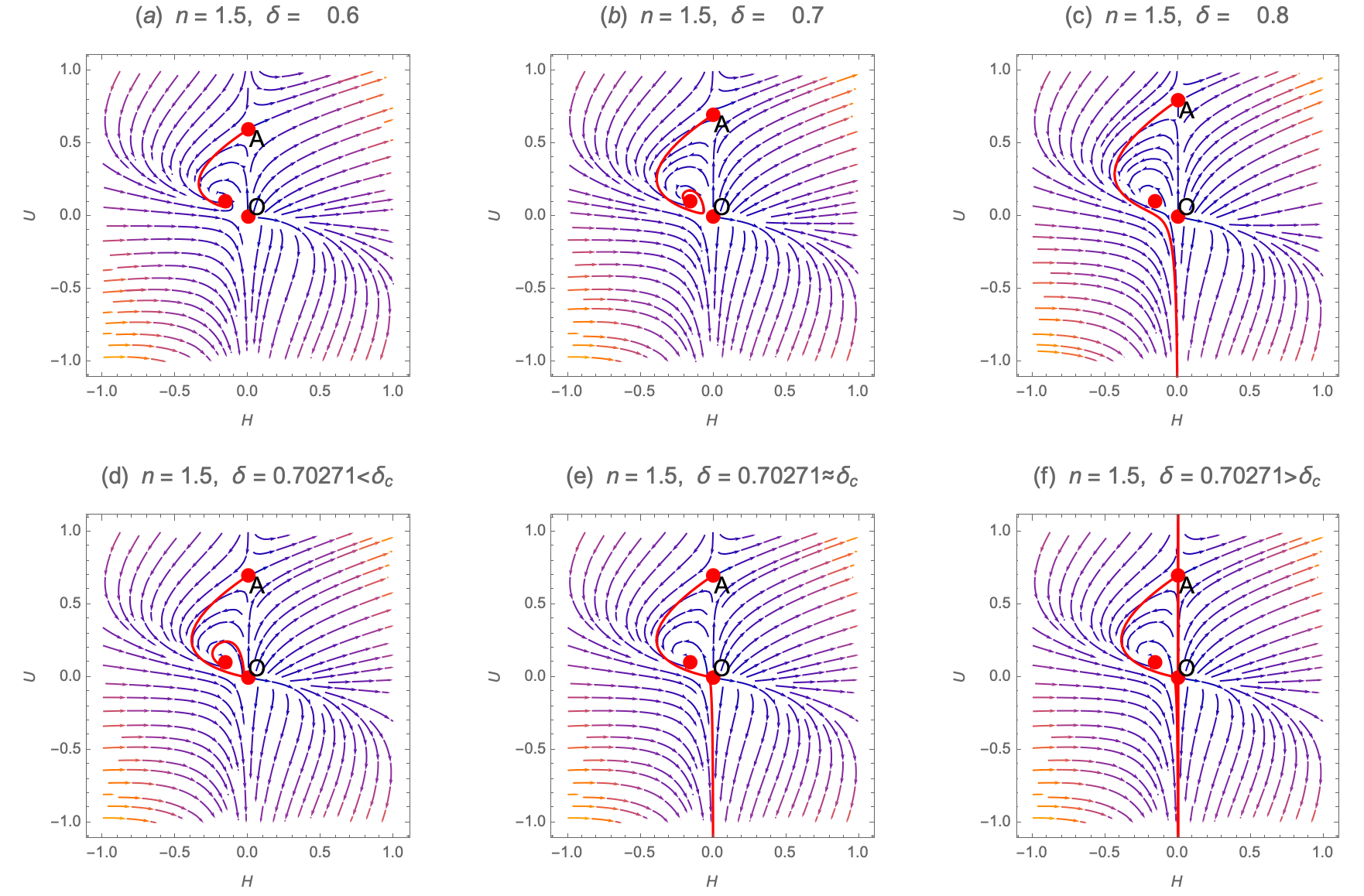}
    \caption{To demonstrate the bisection method used to find $\delta_c$ for artbitrary $n$, we chose $n=1.5$ as an example of fluid flowing towards the origin. The critical points illustrated in this example are also present in different $n$, except the positions are scaled accordingly. The vector field represents the local variation of $(H,U)$, and the red line represents the integration path. Path A to O represents the inflow, so we change $\delta$ until the path connecting the two points appears.  The upper half of the graph shows the behaviour of the integration path where $\delta$ is much higher or lower than the critical value $\delta_c$. The lower half is plotted with $\delta\rightarrow \delta_c$, which shows high sensitivity to change in $\delta$. The path either enters a limit cycle around point B if $\delta<\delta_c$ or diverges towards infinity if $\delta>\delta_c$.}
    \label{fig:bi}
\end{figure}
The integral paths are then created using the built-in \verb|Mathematica| function \verb|NDSolve| with an initial perturbation of the order $10^{-3}$ along $\mathbf{e}_1$. The value of $\delta$ at different values of $n$ is then found by using the bisection method until the integration path passes the vicinity of O. Figure \ref{fig:bi} shows part of the bisection method for $n=0.3$, where we see a small difference in $\delta$ causes the trajectory to either be attracted towards B or shoot off to infinity.

The instability shown in figure \ref{fig:bi} is due to the imperfection of computational perturbation since $\delta\mathbf{r}$ is not infinitesimal. When $\delta<\delta_c$, the cycling path to B corresponds to an oscillation between $U$ and $H$, while $\delta>\delta_c$ shows that the fluid plunges into a sink at a finite distance. Both are divergent behaviours as the fluid converges and self-similarity breaks down (Gratton \& Minotti [1990]).

\begin{figure}
     \centering
     \begin{subfigure}{0.45\textwidth}
         \centering
         \includegraphics[width=\linewidth]{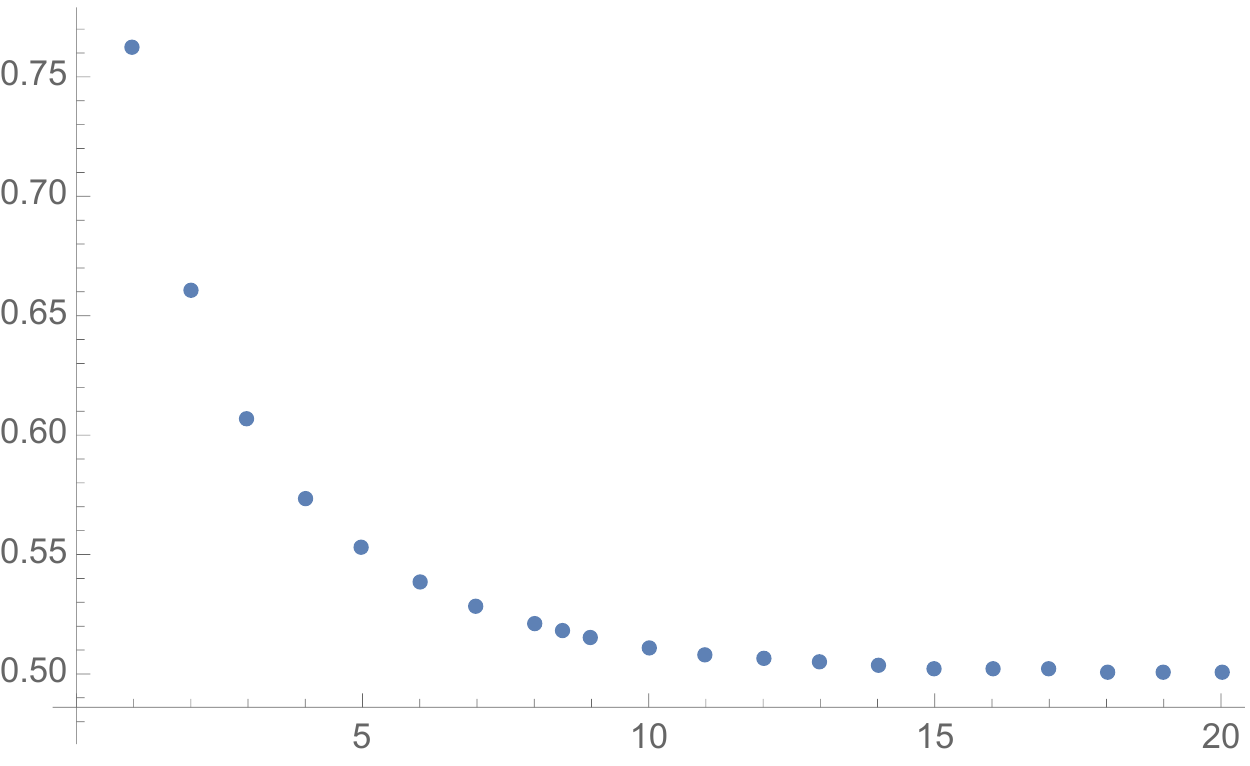}
         \caption{$\delta_c$ against $n$}
         \label{fig:normal}
     \end{subfigure}%
     \begin{subfigure}{0.45\textwidth}
         \centering
         \includegraphics[width=\linewidth]{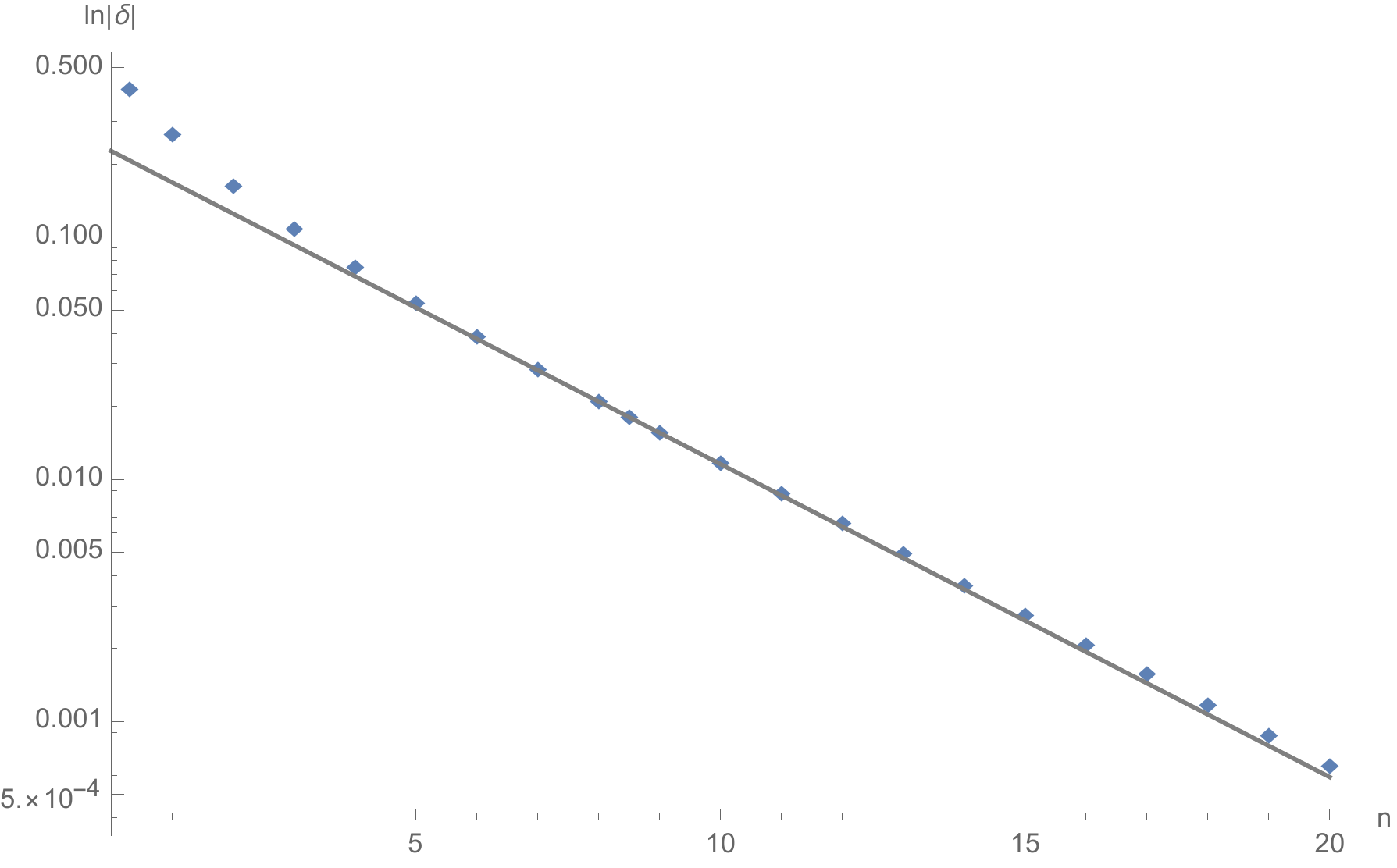}
         \caption{$ln|\delta_c-0.5|$ against $n$}
         \label{fig:log}
     \end{subfigure}
    \caption{ \subref{fig:normal} The critical value of the self-similarity parameter $\delta_c$ for different power-law channels $n$. The data shows a clear trend of asymptotic behaviour approaching $1/2$ as $n$ increases. \subref{fig:log} A linear best-fit of $a\exp(-kn)+1/2$ together with the $(n,\delta)$ data points on the logarithmic scale. The values of $a$ and $k$ were found computationally. There is a close agreement between the data and the linear trend line in the region of large $n$, and the model breaks down as $n\rightarrow 0$.}
    \label{fig:exp}
\end{figure}

\begin{figure}
    \centering
    \includegraphics[width=0.8\linewidth]{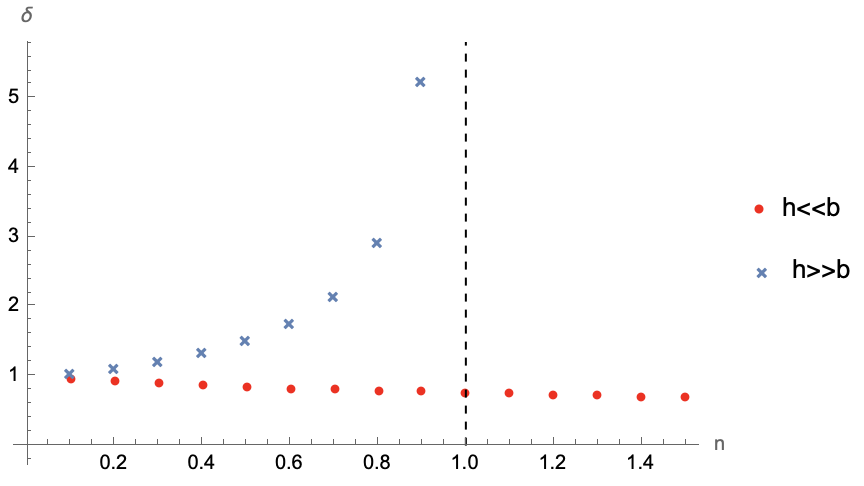}
    \caption{Comparison between the similarity parameter $\delta$ when $h\gg b$ and $h\ll b$. The $h\gg b$ case studied by Zheng et al [2014] shows a diverging trend approaching $n\rightarrow 1^-$, while the $h\ll b$ case has similarity solution of the second kind for all $n$. Importantly, $\delta$ agrees for both cases at $n=0$.}
    \label{fig:withZCS}
\end{figure}

Figure \ref{fig:exp} shows the $\delta_c$ value obtained for various $n$ and the logarithmic plot, which proves to be approximately linear except in the region $n\rightarrow 0$. The $\delta(n)$ then decays exponentially towards $\delta_\infty=0.5$. Fitting the equation as
\begin{equation}
    \delta(n) = ae^{-kn}+\frac{1}{2},
\end{equation}
and using the \verb|FindFit| function, we find
\begin{equation}
    \begin{matrix}
        a\approx &0.23,\\
        k\approx &0.30.
    \end{matrix}
\end{equation}
This result suggests similarity solutions of the second kind exist for all $n$. Specifically, as $n\rightarrow \infty$, the geometry of the channel approximates to an infinite wall with axisymmetric current flowing towards the origin, which agrees with the model as $\delta\rightarrow 1/2$.
A comparison with the results of the "deep" ($h\gg b$) case studied by Zheng et al [2014] is presented in figure \ref{fig:withZCS}, which shows divergence behaviour as $n\rightarrow 1^-$ in the deep case, in contrast to no divergence in the shallow case.
\section{Summary}
We have investigated shallow, viscous flow in a power-law channel, considering both inflow and outflow. We showed that outflow is described after some time by the similarity solution of the first kind and evaluated how long it takes for this similarity solution to be of a required accuracy. We found similarity solutions of the second kind for inflow in different channels using the bisection method. The similarity solutions of the second kind is determined by the similarity variable $\delta$, which decays exponentially as the power of the channel increases. The parameter there exhibits no divergent behaviour, contrary to the behaviour of the solutions in Zheng [2014] and the subsequent review paper [2022].
\section*{Acknowledgement}
We thank H.A. Stone and Z. Zheng for inspiring the idea for this project and T.V. Ball and J. Webber for their numerical simulations. M-S.L. thanks H.E.H., who supervised and hosted this project and Homerton College, Cambridge for their financial support through the Victoria-Brahm-Schild Scholarship. For inspiration and friendship, M-S. L. is grateful for A. Cox, M. Loncar, M. Roach, J. Saville, and O.S. Wilson.
\appendix
\section{Outflow boundary condition}
\label{appendix:bc}
Numerically solving the outflow problem requires the solution to satisfy both the PDE,
\begin{equation}
    \frac{\partial h}{\partial t} -\frac{\beta}{x^n}\frac{\partial}{\partial x}(x^nh^3\frac{\partial h}{\partial x}) = 0,
    \label{eq:height2}
\end{equation}
and the conservation condition ($\alpha=0$),
\begin{equation}
    \int_0^{x_f(t)}h(t,x)x^ndx=B,
    \label{eq:conservation2}
\end{equation}
at all times. This proves to be an issue because not only does the boundary condition change in time $x_f=x_f(t)$, an integral boundary condition is not as easy to discretise as differential boundary conditions. We can find appropriate differential boundary conditions corresponding to the integral conservation by considering the integral of equation \eqref{eq:height2} over the simulation domain $x\in(0,L]$ after multiplying $x^n$ on both sides,
\begin{equation}
    \int_0^Lx^n\frac{\partial h}{\partial t}dx=\frac{\partial}{\partial t}\int_0^Lx^nh(t,x)dx=\beta \Big[x^nh^3\frac{\partial h}{\partial x}\Big]^L_0,
    \label{eq:appendix1}
\end{equation}
where the open lower bound is to avoid dividing by zero. In the actual simulation we set the bounds to be $x\in[l,L]$, where $l>0$ and is close to zero. We can further enforce that $h(t,x)\approx 0$ for $x> x_f(t)$ because $(0,L]$ covers the activity of the fluid ($L>x_f$). Condition \eqref{eq:conservation2} leads to
\begin{equation}
    \int_0^{x_f(t)}x^nh(t,x)dx=\int_0^{L}x^nh(t,x)dx=B.
\end{equation}
Substituting into the last result \eqref{eq:appendix1}, we obtain
\begin{equation}
    \beta L^nh^3\frac{\partial h}{\partial x}\Big|_{x=L}-\beta x^nh^3\frac{\partial h}{\partial x}\Big|_{x\rightarrow 0}=0.
\end{equation}
Physically, both terms are negative, so must vanish,
\begin{subequations}
    \begin{equation}
        \frac{\partial h}{\partial x}\Big|_{x=L}=0,
    \end{equation}
    \begin{equation}
        x^nh^3\frac{\partial h}{\partial x}\Big|_{x\rightarrow 0}=0.
    \end{equation}
\end{subequations}
This method provides us with an easier way of dealing with the conservation condition.
\section{Table of result of $\chi(n)$}
\label{appendix:table}
The table below displays $\chi$ calculated computationally for different values of $n$. 
\\

\begin{center}
\def~{\hphantom{0}}
    \begin{tabular}{lc}
        n	& $\chi(n)$\\[3pt]
        1.0 &	0.09828 \\
        1.5 &	0.07337\\
        2.0 &	0.05649\\
        2.5 &	0.04571\\
        3.0 &	0.03719 \\
        3.5 &	0.03164\\
        4.0 &	0.02639 \\
        4.5 &	0.02321 \\
        5.0 &	0.02050 \\
        5.5 &	0.01824 \\
    \end{tabular}\quad
    \begin{tabular}{lc}
        n	& $\chi(n)$\\[3pt]
        6.0 &	0.01580\\
        6.5 &	0.01432 \\
        7.0 &	0.01275 \\
        7.5 &	0.01150 \\
        8.0 &	0.01036\\
        8.5 &	0.01000\\
        9.0 &	0.00875 \\
        9.5 &	0.00800 \\
        10.0 &	0.00756 \\
        10.5&   0.00697\\
    \end{tabular}
\end{center}

\label{table:table}

\nocite{*}
\bibliographystyle{jfm}
\bibliography{jfm-instructions}

\end{document}